\documentclass[aip,rsi,preprint,graphicx]{revtex4-1} 
\usepackage{amsmath}
\usepackage{graphicx}
\usepackage[section]{placeins}
\usepackage{amsmath}
\usepackage{color}

\draft 

\begin{document}
   
\title{Specific heat measurement of thin suspended SiN membrane from 8~K to 300~K using the 3$\omega$-V$\ddot{o}$lklein method}

\author{Hossein Ftouni} \author{Dimitri Tainoff} \author{Jacques Richard} \author{Kunal Lulla} \author{Jean Guidi} \author{Eddy Collin} \author{Olivier Bourgeois}

\affiliation{Institut N\'EEL, CNRS-UJF, 25 avenue des Martyrs, 38042 Grenoble Cedex 9, France}

\date{\today}

\begin{abstract}
We present a specific heat measurement technique adapted to thin or very thin suspended membranes from low temperature (8~K) to 300~K. The presented device allows the measurement of the heat capacity of a 70~ng silicon nitride membrane (50 or 100~nm thick), corresponding to a heat capacity of 1.4x10$^{-10}$~J/K at 8~K and 5.1x10$^{-8}$~J/K at 300~K. 
Measurements are performed using the 3$\omega$ method coupled to the V$\ddot{o}$lklein geometry. This configuration allows the measurement of both specific heat and thermal conductivity within the same experiment. A transducer (heater/thermometer) is used to create an oscillation of the heat flux on the membrane; the voltage oscillation appearing at the third harmonic which contains the thermal information is measured using a Wheatstone bridge set-up.
The heat capacity measurement is performed by measuring the variation of the 3$\omega$ voltage over a wide frequency range and by fitting the experimental data using a thermal model adapted to the heat transfer across the membrane. The experimental data are compared to a regular Debye model; the specific heat exhibits features commonly seen for glasses at low temperature.
\end{abstract}


\maketitle 

\section{Introduction}

The study of thermal phenomena at the nanoscale has received great attention in the recent years due to the remarkable properties that differ significantly from their bulk counterparts \cite{A.D.McConnell2005}.
When the microstructural length scales of a material become comparable to the mean free path of the phonon,
surfaces start to influence the overall thermal transport \cite{Cahill2003,Kim2006}.
These specific thermal properties can be of great use in various applications such as 
thermoelectricity or more generally energy conversion devices \cite{Hicks1993,Shakouri2011}. Apart from the technological considerations, the study of thermal properties at the nanoscale presents fundamental questions related to the interaction of heat transfer and microstructure at these small length scales: effect on the mean free path \cite{prbheron,condmat}, on the dispersion relations and then on the average phonon group velocities etc... \cite{cuffe}. These effects are also of great interest in amorphous materials like silicon nitride, a material having large potential device applications due to its low thermal conductivity \cite{Zink2004,Revaz2005,Pohl2002,zink2013}. 

Accurately tailoring the thermal properties of nanoscale systems requires the fabrication of very small samples. Hence, innovative techniques able to measure accurately the reduced values of the thermal properties are a growing need. The measurement of thermal transport properties in thin films has been improved significantly in recent years.  

The 3$\omega$ method is generally used to measure the thermal conductivity of semi-infinite materials \cite{F1990}. It has been also shown in the past that this technique based on a dynamic measurement can be used to extract the specific heat \cite{Birge1987}. However, this has never been demonstrated on a membrane system.

Here by coupling the 3$\omega$ method \cite{Cahill1990} and the V$\ddot{o}$lklein geometry \cite{Voelklein2013} (elongated suspended membrane), we present a system designed to measure the specific
heat of silicon nitride suspended membrane (50~nm and 100~nm thick) over a wide temperature range (8 to 300~K). This work is an extension of the recently proposed device to measure the thermal conductivity \cite{jain,Sikora2012,Sikora2013}. The major advantage of the proposed technique comes from the concomitant measurement of the two important thermal properties of materials: the thermal conductivity ($k$) and the specific heat ($c$) by the measurement at different thermal excitation frequencies on the same sample; low frequency for $k$ and high frequency for $c$. Although less sensitive, this technique offers possibilities that cannot be obtained easily from classical specific heat measurement like ac calorimetry \cite{bourg1,Lopeandia2010,rydh}, fast scanning calorimetry \cite{viejo} or relaxation calorimetry \cite{Revaz2005}.

\section{Experimental}
The principle of the method consists in creating a sinusoidal Joule heating generated by an a.c. electrical current of frequency $\omega$
across a transducer centered along the long axis of a rectangular membrane. The center of
the membrane is thermally isolated from the frame and
hence its temperature is free to increase. The temperature oscillation of the membrane is at 2$\omega$ and is
directly related to its thermal properties by its amplitude and frequency respectively. By measuring the $V_{3\omega}$
voltage appearing across the transducer, it is possible to
deduce the thermal conductivity and the specific heat of the membrane.
The transducer is made out of a material whose resistance is strongly temperature dependant. It serves as a
thermometer and heater at the same time.

\begin{figure}
\begin{center}
 \includegraphics[width=8cm]{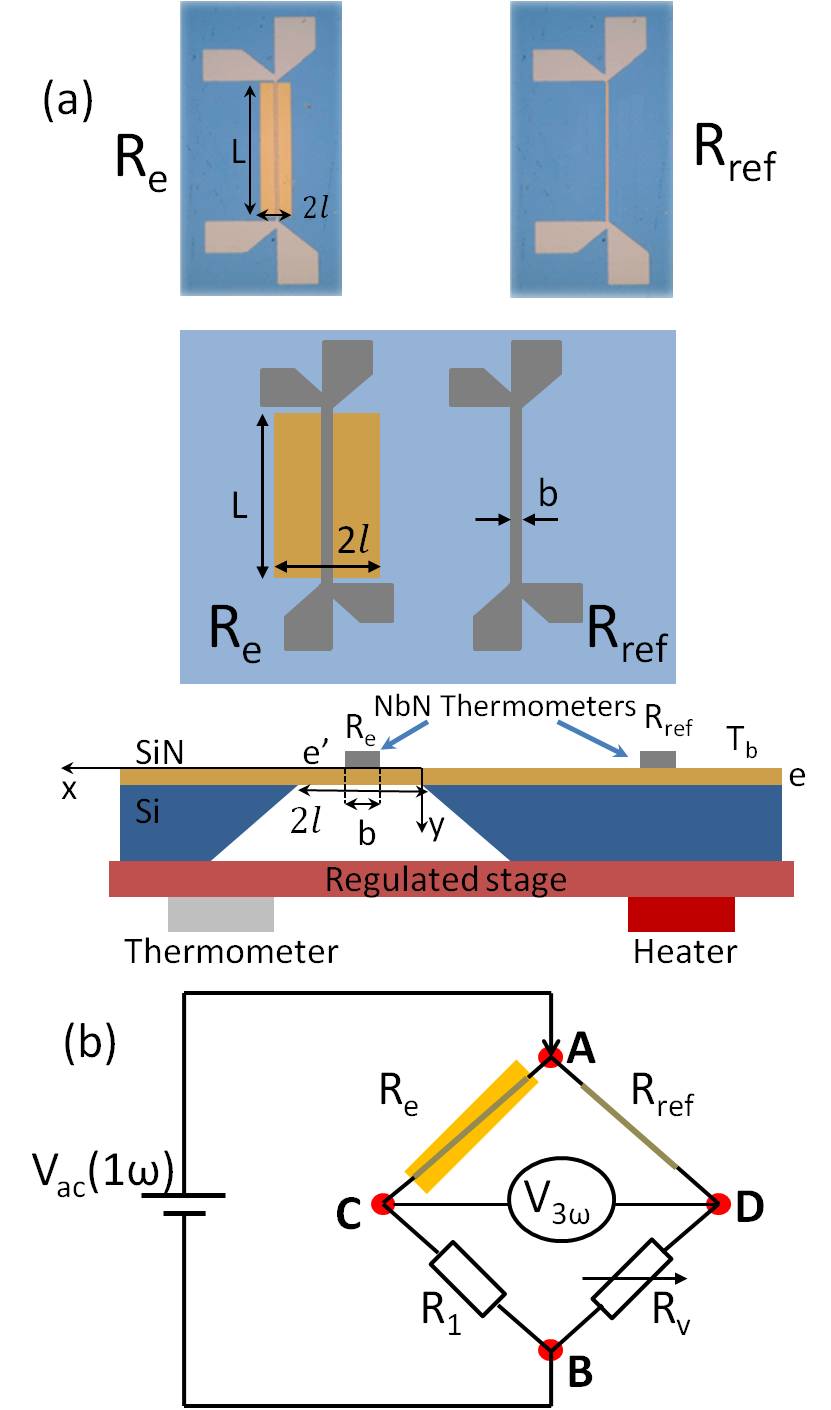}
 \end{center} 
 \caption{(a) Photograph of the two NbN thermometers deposited on the membrane and on the bulk region; below, the schematic of the measurement device installed on the controlled temperature stage. (b) Electrical schematic of the Wheatstone bridge.}
 \label{thermometre}
\end{figure}

However, the measured voltage across the transducer include the $V_{1\omega}$ ohmic component which is usually larger than the $V_{3\omega}$ thermal component by a factor of 10${^3}$. 

By using a specific Wheatstone bridge \cite{Birge1987} we strongly reduce the component
of the measured voltage at angular frequency 1$\omega$. The
bridge consists of the measured sample with its respective resistor $R_{e}$, which
is the niobium nitride (NbN) thermometer on the SiN membrane, the reference
thermometer $R_{ref}$ deposited on the bulk region of the chip which has the same geometry and deposited in the same run as the
transducer on the membrane (see Fig. \ref{thermometre}), an adjustable resistor $R_{v}$, and an equivalent
nonadjustable resistor $R_{1}$ = 50~kOhm. The two resistors $R_{v}$ and $R_{1}$ are
positioned outside the cryogenic system. (see more details about the electrical set-up in the reference \cite{Sikora2012}$^{,}$ \cite{Sikora2013}). 
Since the reference thermometer is not on the membrane, its temperature
remains at the bath temperature $T_{b}$ and therefore, its resistance does not change.
The elevation of temperature due to self-heating of the reference
transducer is neglected thanks to the thermal contact to the quasi-infinite reservoir of
the bulk silicon.

In that geometry, the voltage at 1$\omega$ has been reduced by a factor of
10${^3}$. Thus, it is possible to measure the $V_{3\omega}$ signal with a high sensitivity on the Wheatstone bridge output without the 1$\omega$ component saturating the dynamic reserve of the lock-in amplifier. 

The two NbN thermometers (Nb$_1$N$_{1.7}$) have practically the same temperature behavior as they have
been deposited simultaneously on the SiN substrate. However,
due to the presence of inhomogeneity in the deposition process,
there is a slight difference of resistance. Thus, the $R_{v}$
resistor is used to balance the bridge. Thanks to the Wheatstone
bridge, the $V_{3\omega}$ signal is larger than the $V_{1\omega}$ signal.

The geometry of the membrane measured in this study is the following: 300$\mu$m wide and 1.5~mm long; the transducers that are patterned using regular clean room processes are centered along the main axis and have various widths: 5, 20, 30 and 40$\mu$m. They are made of NbN thin film
that are grown using a dc-pulsed magnetron sputtering from a high purity niobium target in a gas mixture of argon and nitrogen. This type of high sensitivity thermometer is described in detail in Ref. \cite{Bourgeois2006} Its temperature coefficient of resistance (TCR) can be tailored over a wide temperature range, from low temperature \cite{Heron2009a} to high temperature \cite{Lopeandia2010}. Hence, depending on the stoichiometry, the electrical properties of the NbN can vary a lot. For the SiN measurement, the thermometer has been designed for the 10~K to 320~K temperature range. Typically, the resistance of the thermometer is about 100~kOhm at room temperature with a TCR of 10$^{-2}$ K$^{-1}$ and 1~MOhm at 70~K with a TCR of 0.1~K$^{-1}$. The resistance of the thermometer on the membrane is calibrated using a standard four probe technique between 4~K and 330~K in a $^{4}$He cryostat. The devices in the cryogenic vacuum are protected by a thermal copper shield maintained at the $T_{b}$ temperature to reduce the heat radiation and installed on a temperature regulated stage as schematized in the Fig. \ref{thermometre}. The stage temperature is regulated with a stability of the order of few milliKelvin. The stage temperature $T_{b}$ may be varied from 4~K to more than 330~K.

\section{General solution of heat flow}

The NbN thermometer is calibrated in a four-probe configuration (see Fig. \ref{thermometre}). The two outside contacts are used to apply
an ac current while the voltage is measured by the two inside contacts. As the membrane is suspended, its temperature is free to fluctuate. The specimen is maintained in a high vacuum and the whole setup is heat shielded to the substrate temperature to minimize heat losses through gas convection and radiation. Thus, in such configuration and with an ac electrical current of the form $I_{0}$sin($\omega$t) passing through the specimen, the 1D partial differential equation of the heat flux across the membrane is given by:

\begin{equation}
\frac{\partial^{2}T(x,t)}{\partial x^{2}}=\frac{1}{D}\frac{\partial T(x,t)}{\partial t}
\label{general-eq}
\end{equation}

with the initial and boundary conditions:\\

$\left\{ 
\begin{array}{ll}
T(x,t=0)=T_{b} \\
T(x=0,t)=T_{b} \\
C'(T)\frac{\partial T(x,t)}{\partial t}_{x=\ell}=P(t)-eLk\frac{\partial T}{\partial x}_{x=\ell}
\end{array} \right.$
\\

\begin{align}
&\rm{where} &C'=&\rho_{NbN}c_{NbN}Le'\frac{b}{2}+c\rho\frac{b}{2}Le \
\end{align}

with $c$ the specific heat, $\rho$ the density, $D=k/\rho c$ the diffusivity of the SiN membrane, and $k$ the thermal conductivity. 
The total dissipated power $P(t)$ is used to heat both the thermometer and the part of the membrane under the thermometer, and the rest of the membrane:

\begin{figure}
\begin{center}
 \includegraphics[width=8cm]{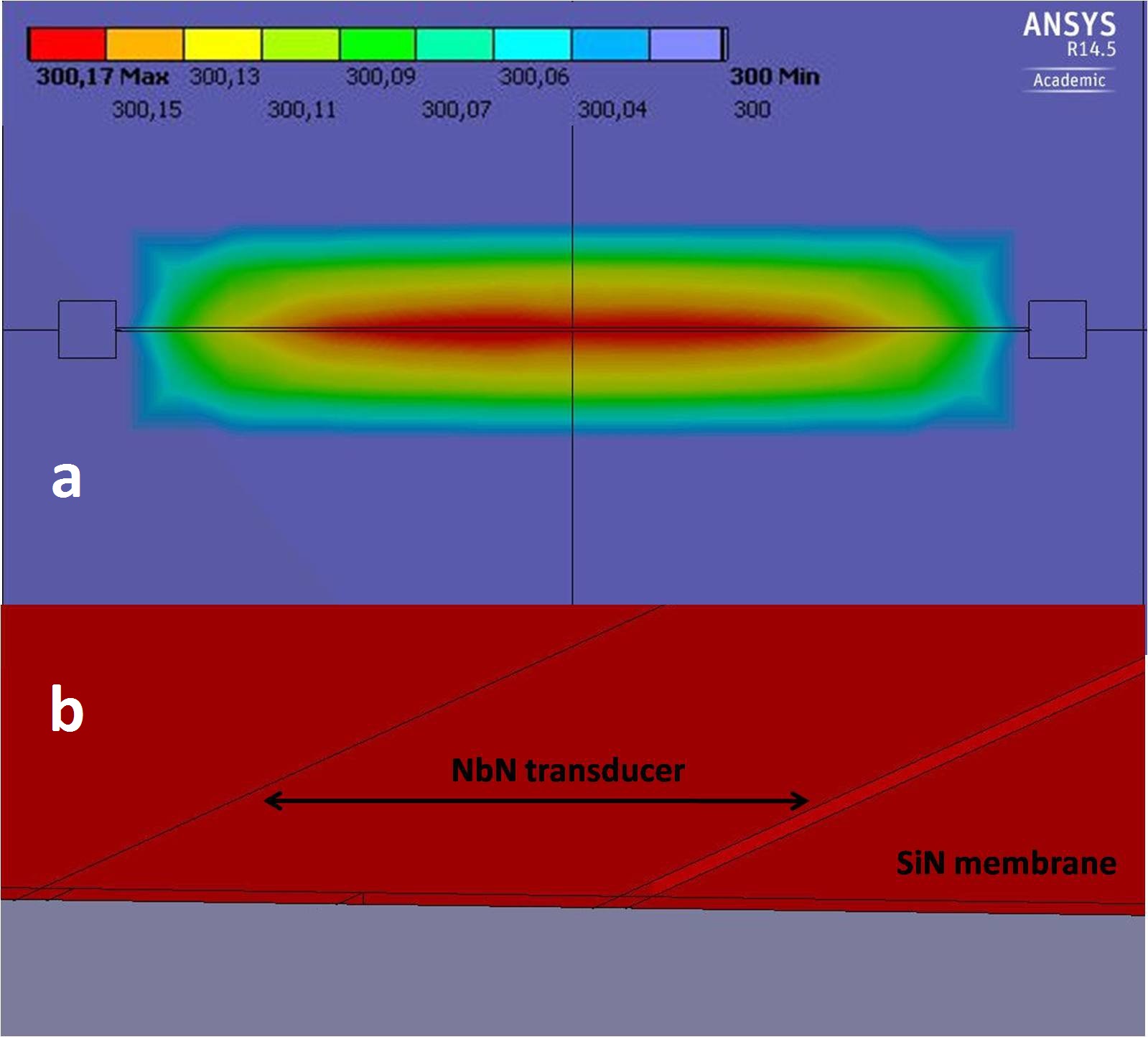}
 \end{center} 
 \caption{ Finite element simulation of a 100~nm thick SiN membrane (width 300~$\mu$m and length 1.5~mm) including NbN transducer: (a) Top view of the membrane. The isotherm lines lie along the transducer except at the edge of the membrane;(b) temperature profile of the cross-section of the membrane. The NbN transducer (width 5~$\mu$m) and the membrane are at the same temperature; this temperature is constant over the entire thickness of the membrane.} 
 \label{SimulF}
\end{figure}

The general solution of Eq. \ref{general-eq} is:

\begin{equation}
T(x,t)=\frac{P_{0}sh\left[\omega'(1+j)x\right]e^{j2\omega t}}{(1+j)Sk\omega'ch\left[\omega'(1+j)\ell\right]+j2C'\omega sh\left[w'(1+j)\ell\right]}
\label{Tac}
\end{equation}
with $\omega'=\sqrt{\frac{\omega}{D}}$, $S$=$eL$, $P_{0}$=$\frac{RI_{0}^{2}}{4}$ and $I=I_{0}sin(\omega t)$.
We can also write Eq. \ref{Tac} using exponential notation:
\begin{equation}
T(x,t)=\frac{P_{0}}{D_{0}^{1/2}}\left[sin^{2}(\omega'x)+sh^{2}(\omega'x)\right]^{1/2}e^{j2\omega t+\varphi}
\label{tx}
\end{equation}
with $\varphi$ the phase and $D_{0}^{1/2}$ the absolute value of the denominator of Eq. \ref{Tac}.
After development in Taylor expansion to first order in $\omega$, the expression of the absolute value of the temperature  $T_{m}(\ell)$ can be written as follows:
\begin{equation}
T_{m}(\ell)=\frac{P_{0}}{K_{p}{\left[1+\omega^{2}\left(4\tau^{2}+\frac{2\ell^{4}}{3D^{2}}+\frac{4\tau \ell^{2}}{3D}\right)\right]}^{1/2}}
\end{equation}
with $K_{p}$=$\frac{kS}{\ell}$, $\tau$=$\frac{C'}{K_{p}}$ and $D$ the thermal diffusivity.
The general form of the voltage across the thermometer, $V_{AC}(\omega)$ is given by :

$\begin{array}{ll}
V_{AC}(\omega)=R^{'}_{e}I, \rm{where} 
\\
R^{'}_{e}= R_{e}\left[1+\alpha\left|T\left(l,t\right)\right| cos\left(2\omega t+ \varphi\right)\right],
\\
I=\frac{V_{ac}e^{j\omega t}}{\left(R_{1}+R_{e}\right)}
\\
\end{array}$

with $I$ the current flowing through the thermometer. 
Then, the general expression of the voltage, between A and C, can be written as follows:
\begin{equation}
V^{rms}_{AC}(\omega)=\frac{V^{rms}_{ac} R_{e}\left[1+\alpha\left|T\left(l,t\right)\right| cos\left(2\omega t+ \varphi\right)\right] cos(\omega t )}{([R_{1}+R_{e}\left[1+\alpha\left|T\left(l,t\right)\right| cos\left(2\omega t+ \varphi\right)\right]^{2})^{1/2}}
\end{equation}
with $V_{ac}$ the voltage put on the Wheatstone bridge (between A and B), $\varphi$ the thermal phase.

The absolute value of $V_{3\omega}$ is given by:
\begin{equation}
\left|V_{3\omega}^{rms}(\omega)\right|= \frac{V^{rms}_{ac}\alpha R{e}R{_1}\left|T(l,t)\right|}{2(R_e + R_1){^2}  }
\end{equation}
and the phase by:
\begin{equation}
\left\{ 
\begin{array}{ll}
\varphi_{V_{3\omega}}(\omega)= \varphi
= \varphi_{1} + \varphi_{2} \\
tg\varphi_{1}= -\frac{\omega^{'}\left[ch(\omega^{'}l)cos(\omega^{'}l)+sh(\omega^{'}l)sin(\omega^{'}l)\right]+\omega \tau sh(\omega^{'}l)cos(\omega^{'}l)}{\omega^{'}\left[ch(\omega^{'}l)cos(\omega^{'}l)-sh(\omega^{'}l)sin(\omega^{'}l)\right]-\omega \tau ch(\omega^{'}l)sin(\omega^{'}l)}\\
tg\varphi_{2}= \frac{tg(\omega^{'}l)}{th(\omega^{'}l)}
\end{array} \right.
\label{phase3w1}
\end{equation}
\\
Then the general expression of $V_{3\omega}$ becomes:
\begin{equation}
\left|V_{3\omega}^{rms}(\omega)\right|=\frac{\alpha   (V_{ac}^{rms})^{3}R_{1}R_{e}^{2}}{4K_{p}\left(R_{e}+R_{1}\right)^{4}\left[1+\omega^{2}\left(4\tau^{2}+\frac{2l^{4}}{3D^{2}}+\frac{4\tau l^{2}}{3D}\right)\right]^{1/2}}
\label{V3w} 
\end{equation}
The thermal conductivity can be extracted simultaneously using the same fit. The extracted specific heat values from the 1D and 2D models show very comparable results. A difference of 0.8$\%$ is observed. Thus, in the following the 1D model is used for the sake of simplicity.

Finite element simulations have been performed to confirm this assumption unsing the ANSYS platform \cite{ansys}. Results are displayed on Fig. \ref{SimulF}. Heat flows from the transducer to the quasi-infinite reservoir of the bulk silicon. Except at the edges of the membrane the temperature along the heater is nearly the same. We can also verify on Fig. \ref{SimulF} that the temperature is uniform over the membrane thickness confirming the assumption made for analytical calculations.

\section{Results and discussion}

\subsection{Specific heat measurement} 
\begin{figure}
\begin{center}
 \includegraphics[width=10cm]{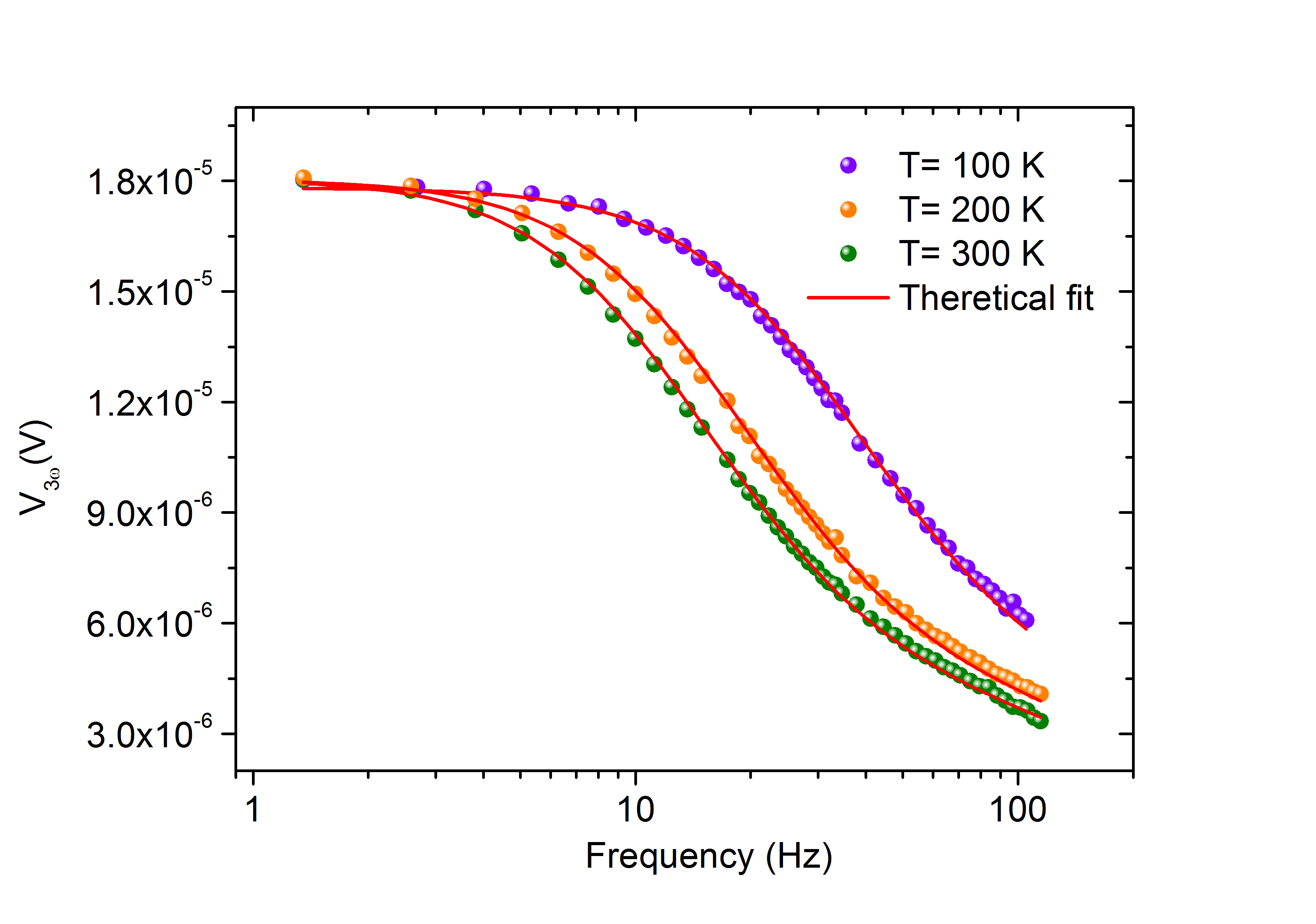}
 \end{center} 
 \caption{ V$_{3\omega}$ voltage measurements versus frequency at different temperatures for a 100~nm thick membrane.}
 \label{V3w(T)}
\end{figure}

At fixed temperature, the specific heat of the membrane is extracted by fitting the 3$\omega$ voltage data versus the frequency using Eq. \ref{V3w}. The thermal cut-off frequency increases when the temperature drops down as shown in Fig. \ref{V3w(T)}. To assure that the frequency dependence of the 3$\omega$ comes only from thermal origin and there are no electrical dependencies, we assume that the two thermometers present an electrical capacitance. By fitting the 1$\omega$ Wheatstone output voltage using the electrical model explained in the previous publication \cite{Sikora2012}$^{,}$ \cite{Sikora2013}, we are able to estimate this capacitance to be around hundred picofarad and then can not affect the thermal frequency cut-off above 1~kHz, which is above the frequency measurement range (1 to 100~Hz). A geometrical effect of the thermometer width on the thermal frequency cut-off was observed, this effect is discussed in detail in the following section.

\subsection{Effect of a finite transducer width}

\begin{figure}[h]
\begin{center}
 \includegraphics[width=10cm]{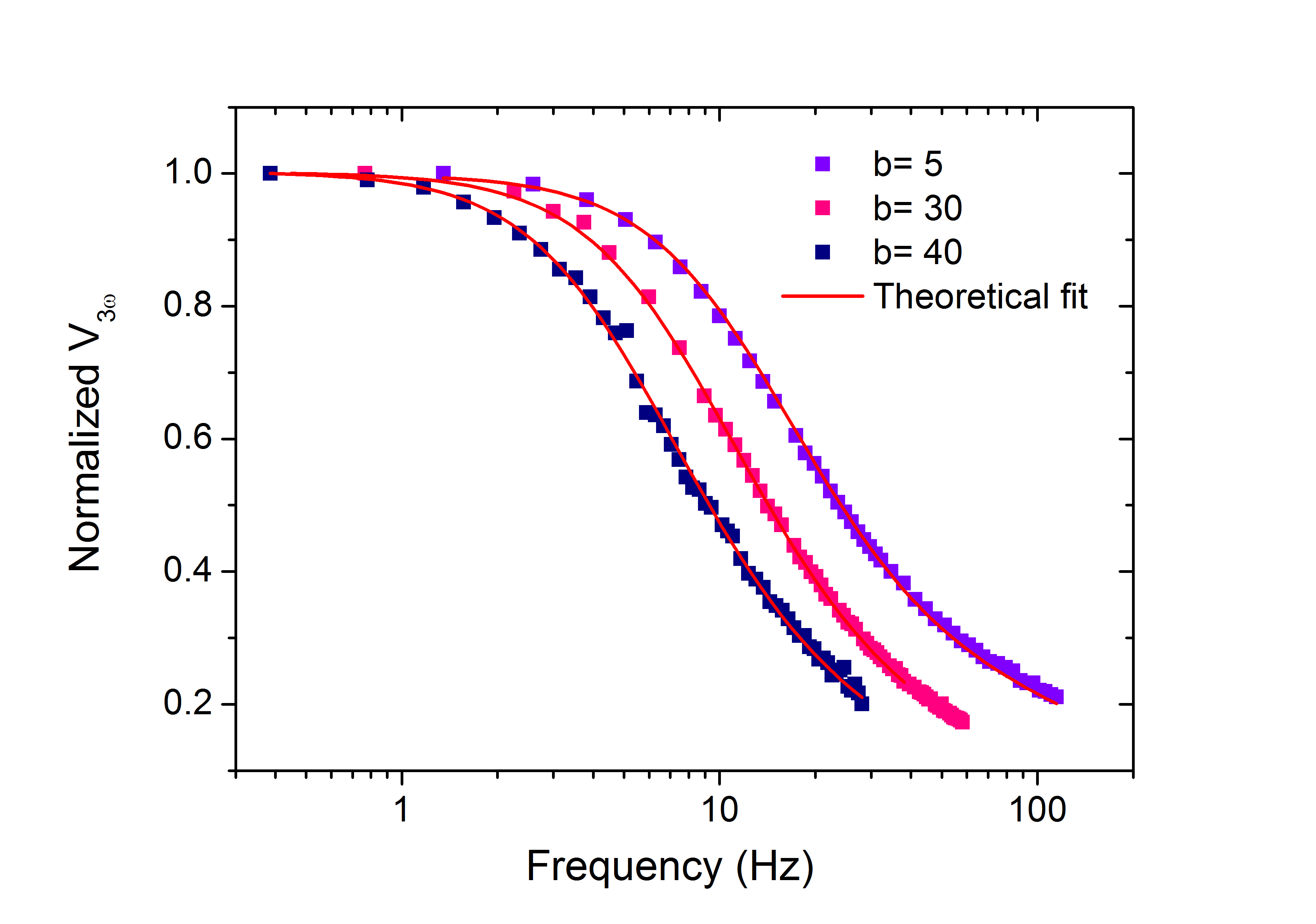}
 \end{center} 
 \caption{V$_{3\omega}$ voltage measurements versus frequency normalized to the low frequency value at 250~K using different thermometer width for a 100~nm thick membrane.}
 \label{V3w(Lth)}
\end{figure}

The effect of a finite transducer width has been studied by the measurement of the 3$\omega$ voltage at fixed temperature using different thermometer width. Different measurements have been performed at the same temperature on four distinct samples exhibiting
a large difference in thermal frequency cut-off of the 3$\omega$ voltage (see Fig. \ref{V3w(Lth)}).  

The thermal properties obtained from the theoretical fit are mentioned in Table. \ref{table:V3w(Lth)}. The length $L$ of the thermometer is 1500 $\mu m$. 

\begin{table}[ht]
\caption{Thermal properties parameters at 250~K obtained from the theoretical fit of the measured 3$\omega$ voltage using different thermometer width b.}
\centering 
\begin{tabular}{c c c c} 
\hline\hline 
Sample$\#$ &\hspace{15pt}    b($\mu m$)     &\hspace{15pt} k(W/(m.K)) &\hspace{15pt} C(J/(g.K))\\ [0.5ex]
\hline 
1 & 5 & 3.190 & 0.698 \\ 
2 & 20 & 2.980 & 0.710 \\
3 & 30 & 3.230 & 0.786 \\  
4 & 40 & 3.470 & 1.155 \\ [1ex]
\hline 
\end{tabular}
\label{table:V3w(Lth)} 
\end{table}

The extracted thermal conductivity values present a small variation when the thermometer width increases. When the thermometer width is multiplied by a factor of eight, the extracted thermal conductivity varies at most by 10~$\%$, which is a weak effect. This effect can be explained by the fact that when the width of the thermometer becomes large as compared to the width of the membrane, a gradient of temperature between the centre and its extremity appears and thus cannot be considered like a finite line oscillating at the same temperature to solve the heat transfer equation.

\begin{figure}
\begin{center}
 \includegraphics[width=10cm]{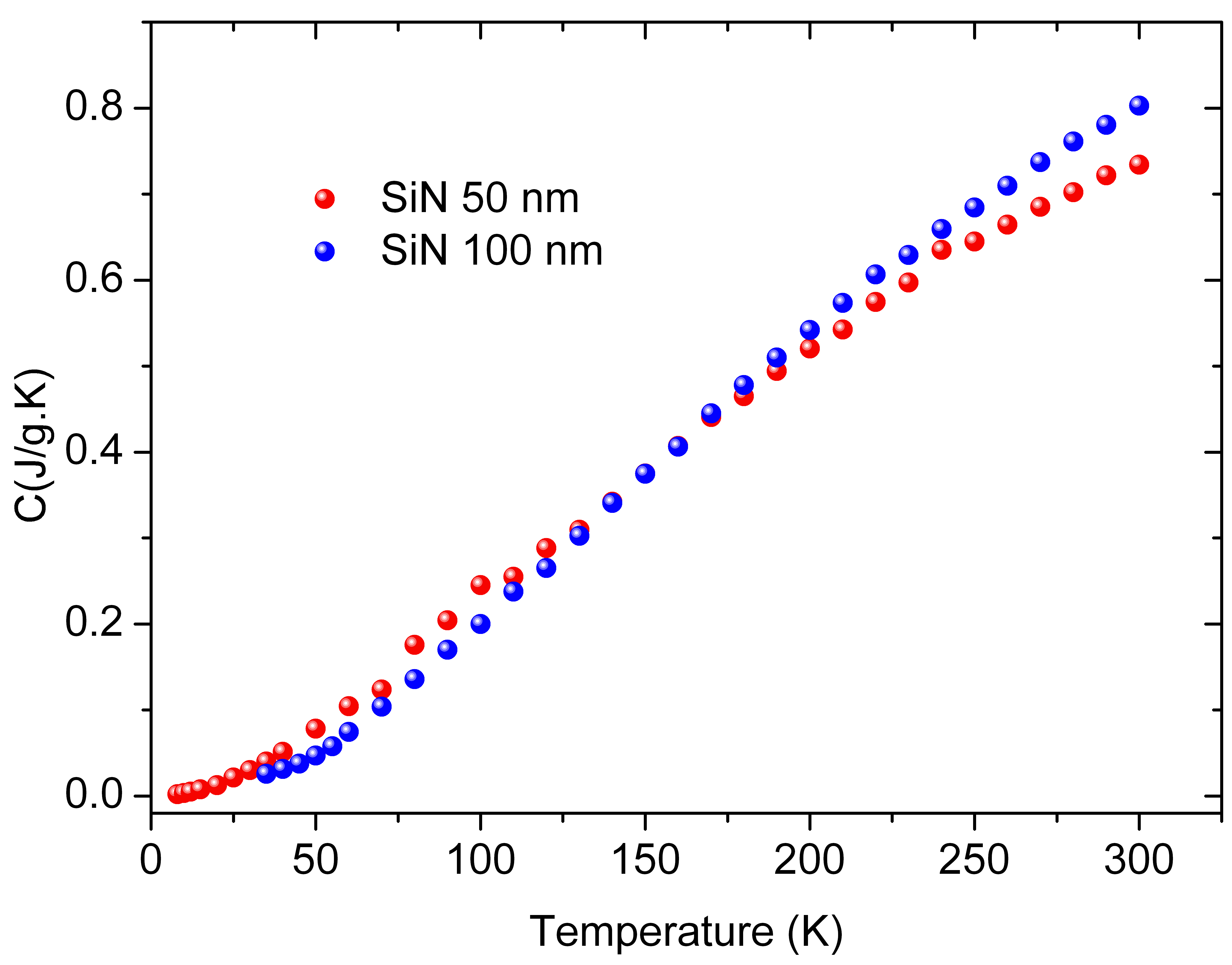}
 \end{center} 
 \caption{The specific heat of the 50~nm and 100~nm thick suspended SiN membranes.}
 \label{C(T)}
\end{figure}

On the other hand, a significant effect of the thermometer width change on the specific heat values was observed. An increase of 65~$\%$ was observed when the thermometer width is multiplied by a factor of eight. This observation is discussed in terms of the thermal penetration depth dependence with the frequency. At low frequency, the entire membrane oscillates at the same temperature with the frequency 2$\omega$ where the thermal penetration depth $\lambda= \sqrt{\frac{2D}{\omega }}$ (with D the thermal diffusivity) is larger than the dimension of the membrane. When the frequency increases, $\lambda$ begins to decrease affecting the overall temperature oscillation of the membrane. At sufficient high frequency, $\lambda$ becomes comparable to thermometer width and then the 3$\omega$ voltage becomes sensitive to the specific heat of the thermometer and the SiN membrane underneath (see Table. \ref{table:V3w(Lth)}). At 100~Hz, the thermal penetration depth is estimated to be around 55~$\mu$m. In the following, the measurements are performed with a thermometer of 5~$\mu$m width; the extracted specific heat values are in perfect agreement with the ones extracted from the experiment done with a thermometer having a width of 20~$\mu$m.
As a conclusion for this part of the study, in order to do a safe experiment, a ratio of at least ten between the width of the membrane and the width of the thermometer has to be respected.

\subsection{Specific heat of the SiN membrane}

\begin{figure}[h]
\begin{center}
 \includegraphics[width=10cm]{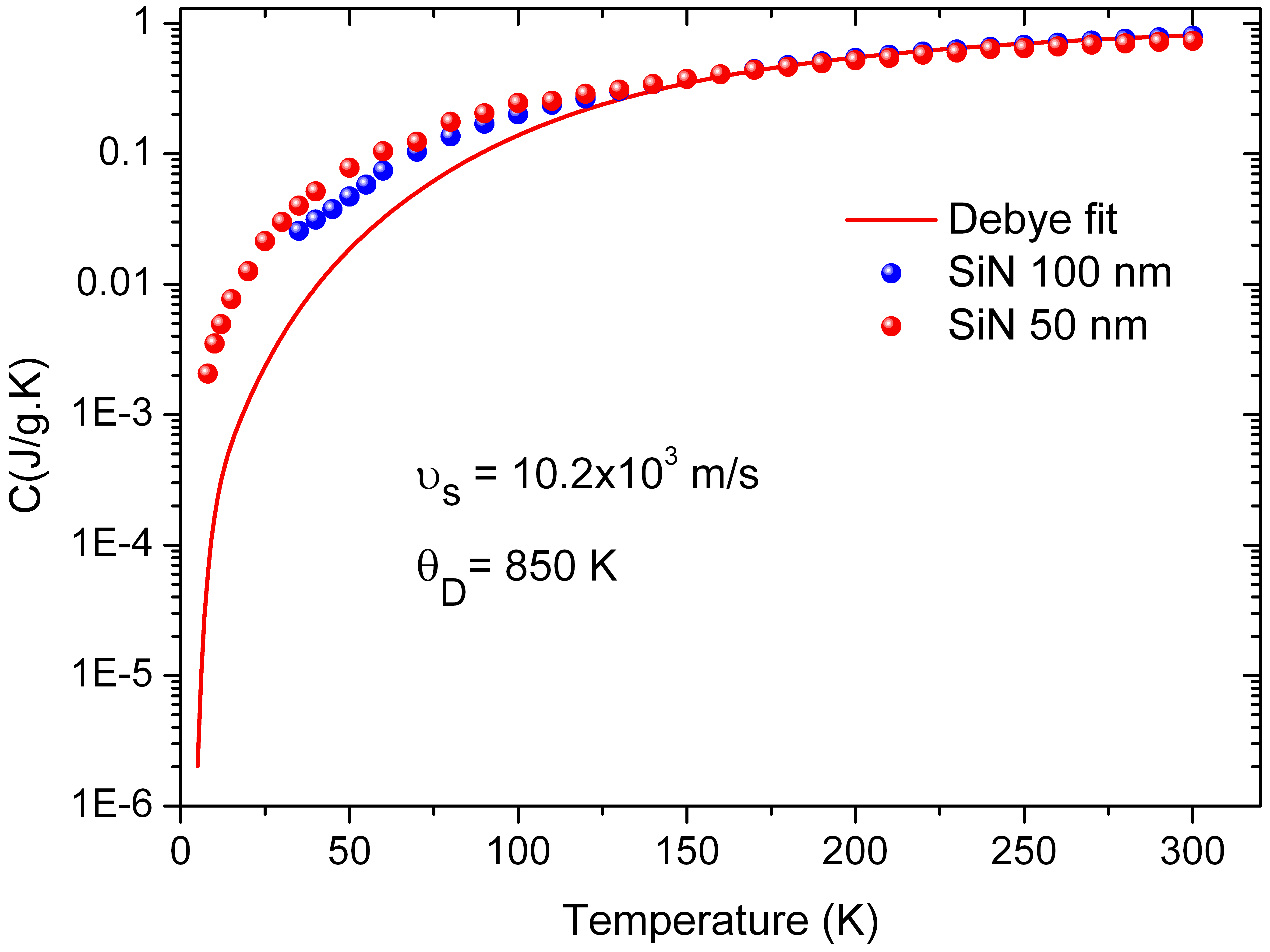}
 \end{center} 
 \caption{The specific heat of a 50~nm and a 100~nm thick SiN membrane and their respective Debye fits on a semi-log plot.}
 \label{Debyefit}
\end{figure}

Fig. \ref{C(T)} and \ref{Debyefit} show the specific heat data of a 50~nm and a 100~nm thick SiN membrane with the corresponding Debye fit plotted versus the temperature. Below 100~K, a deviation from Debye-like specific heat is seen, the specific heat rise is stronger than the Debye $T{^3}$ term as already mentioned for glassy materials but at lower temperature \cite{Gil1993}. 
From the Debye specific heat fit using a sound velocity estimated from a mechanical measurement, the Debye temperature is estimated to be $\theta_{D}$=850~K, a value commonly accepted for amorphous SiN membranes \cite{Zink2004,GUZMAN1976}. As shown in Fig. \ref{C(T)}, only a slight difference between the 50~nm and 100~nm membrane specific heat is observed, illustrating that the reduced dimensions do not affect significantly the specific heat in this temperature range.  

\section{conclusions}  

We have presented measurements of the specific heat of suspended SiN membranes from 8 to 300~K by using the 3$\omega$ method in a V$\ddot{o}$lklein geometry. By fitting the frequency-dependent 3$\omega$ voltage data to Eq. \ref{V3w}, we have obtained the specific heat of the SiN membrane with a sensitivity of 4~nJ/g.K at room temperature. The configuration used for the specific heat measurements also allow the measurement of thermal conductivity of the same sample at low frequency with a very high resolution \cite{Sikora2012,Sikora2013} demonstrating the valuable advantage of this technique; even if the global sensitivity does not reach the performances of competing techniques like ac calorimetry \cite{bourg1,Lopeandia2010,rydh}, fast scanning calorimetry \cite{viejo} or relaxation calorimetry \cite{Revaz2005}.

The Debye temperature has been extracted from the specific heat variation of SiN as a function of temperature. A deviation from Debye $T{^3}$ law has been observed at low temperature as already reported by other authors \cite{Zink2004}. Further measurements down to very low temperature (T$<$10K) are underway. 

This new configuration of the 3$\omega$ method could also be used as a sensor for the measurement of both the specific heat and the thermal conductivity of a given material deposited on the back side of the membrane. This technique provides a platform for the measurement of thermal properties of very thin films, especially for the characterization of transport along the plane.

\section*{acknowledgements}

We acknowledge technical supports from Nanofab, the Cryogenic shop, the Electronic shop and Capthercal from the Institut N\'eel for these experiments. Funding for this project was provided by a grant from La R\'egion Rh\^one-Alpes (CMIRA), by the Agence Nationale de la Recherche (ANR) through the project QNM and by european fundings through the MicroKelvin project and the MERGING project grant agreement No. 309150. We would like to thank P. Gandit, J-E. Lorenzo-Diaz, B. Fernandez, T. Crozes, T. Fournier, E. Andr\'e and J.-L. Garden for help and fruitful scientific exchanges and M. Nunez-Regueiro for financially supporting the ANSYS project via ANR grant TetraFer ANR-09-BLAN-2011.

\end{document}